 \documentclass[final,authoryear,3p,times,twocolumn]{elsarticle}

\usepackage{graphicx}

\usepackage{amssymb}

\journal{Icarus}

\begin{document}

\begin{frontmatter}



\title{Dynamics of escaping Earth ejecta 
  and their collision probability with different Solar System bodies}


\author[l1]{M. Reyes-Ruiz\corref{cor1}}

\author[l1]{C.E. Chavez}

\author[l2]{M.S. Hernandez}

\author[l1]{R. Vazquez}	

\author[l1]{H. Aceves}	

\author[l3]{P.G. Nu\~nez}

\address[l1]{Instituto de Astronom\'{\i}a, 
  Universidad Nacional Aut\'onoma de M\'exico,
  Apdo Postal 877, Ensenada 22800, B.C., M\'exico}

\address[l2]{Facultad de Ciencias, Universidad Autonoma de 
Baja California, Ensenada 22860, B.C., M\'exico}

\address[l3]{Instituto de Estudios Avanzados de Baja California, 
  Ensenada 22800, B.C., M\'exico}

\cortext[cor1]{email: maurey@astro.unam.mx}

\begin{abstract}
It has been suggested that the ejection to interplanetary
space of terrestrial crustal material accelerated in a large impact,
may result in the interchange of biological material
between Earth and other Solar System bodies.
In this paper, we analyze the fate of debris ejected from 
Earth by means of numerical simulations of the 
dynamics of a large collection of
test particles. This allows us to determine the 
probability and conditions for the collision of ejecta with other 
planets of the Solar System. We also estimate the amount of 
particles falling-back to Earth as a function of time after being 
ejected.

The Mercury 6 code is used to compute the dynamics of test particles 
under the gravitational effect of the inner planets in the Solar System 
and Jupiter. A series of simulations are conducted with different ejection 
velocity, considering more than 10$^4$ particles in each case.
We find that in general, the collision rates of Earth ejecta with Venus
and the Moon, as well as the fall-back rates,
are consistent with results reported in the literature. By considering a 
larger number of particles than in all previous calculations we have also 
determined directly the collision probability with Mars and, for the first 
time, computed collision probabilities with Jupiter. We find that 
the collision probability with Mars is greater than values 
determined from collision cross section estimations previously reported.

\end{abstract}

\begin{keyword}

Astrobiology \sep Impact processes \sep Celestial mechanics 


\end{keyword}

\end{frontmatter}



\section{Introduction}

Presently, the collision of kilometer-scale bodies with Earth, 
such as comets or asteroids, is believed to occur on a timescale of the 
order of millions of years (Chapman, 1994). Impacts 
by even greater bodies, with diameter of tens of kilometers, 
such as the Chicxulub event (Kent et al. 1981), are
thought to take place approximately every 10$^8$ years. During the 
Late Heavy Bombardment (LHB) epoch of the Solar System's history, 
both the frequency and diameter of Earth impactors is believed to
be much greater than these estimates (Strom et al. 2005). 

In addition to their catastrophic effect on the diversity of 
life-species
on Earth, giant impacts may also lead to ejecta accelerated with 
velocities greater than the planetary escape velocity, $V_{\rm esc}$. 
Depending on the impactor energy, ejected debris may reach velocities 
significantly higher than $V_{\rm esc}$ reaching Jupiter crossing 
orbits (Gladman et al. 2005). In the impact spallation model of 
Melosh (1984, 1985), material from a thin surface layer of the Earth's 
crust, can be lifted and accelerated to more than escape 
velocity by the interference of impact induced shock waves. Very low
peak-shock pressures are predicted for such ejecta, offering a
plausible explanation of the observed shock levels of meteors of
Lunar and Martian origin.

Once material is ejected to interplanetary space, it will travel in 
orbits that may, depending on the ejection velocity, cross the orbits 
of other planets in the Solar System. Gladman et al. (2005, and references
therein) have analysed the dynamics of such ejecta modelling these 
as a large collection of test particles. They found that,
for low ejection velocities, a fraction of particles 
return to Earth after approximately 5000 years, between 0.6--0.2$\%$ 
for $V_{\infty}$ between 1 and 2 km/s, where $V_{\infty}$ is the velocity 
reached by the particle at very large distances from the Earth. 
An even smaller percentage,
of the order of 0.015\% collides with Venus on a similar timescale. 
In principle, particles ejected with a velocity greater 
than the planet's escape velocity may reach Mars 
or even Jupiter crossing orbits. However, no collisions 
with either body is reported in the calculations 
of Gladman et al. (2005).

It has been suggested that such ejected crustal debris may carry 
along biologic material which may, if it collides with a suitable 
target, serve as seed for the development of life elsewhere in the 
Solar System (Mileikowsky et al. 2000 and Nicholson et al. 2000). 
Additionally, ejected material may return to Earth 
and ''reseed'' terrestrial life after the sterilizing effect of a 
giant impact has passed (Gladman et al. 2005). For this to 
happen, additional constraints are imposed on the time the 
ejecta may remain in space, as well as on the size of the
crustal fragments, as these factors impact the amount of high energy 
radiation to which the biologic material is exposed. 
Wells et al. (2003) have
argued that, with the current characteristics of the space environment
(cosmic rays, x-rays, EUV) exposure for times greater than a few 
thousand years would make biological material nonviable.

In this paper, we analyse the dynamics of particles ejected from 
Earth in a manner similar to the analysis of Gladman et al. (2005)
but improving the statistics by increasing the number of particles 
by more than a factor of three and using a different scheme and code to integrate the 
equations of motion. In section 2 we describe the numerical method 
used, the initial conditions and other details of our simulations.
Results of the various simulations conducted are presented in 
section 3. In section 4 we discuss our results and we present
our concluding remarks in section 5.   


\section{Model description}
\label{sec:problem}

We consider ejecta as test particles moving under the action 
of the gravitational field of the Sun, the Moon, and all planets 
of the Solar System. No other forces are considered in 
the present study. Particles are assumed not to collide
with each other, but may impact any of the massive 
bodies. The dynamics of the planets and test particles system 
is calculated using the Mercury 6.2 code developed by 
Chambers (1999). The code offers several integrator choices, including 
a hybrid integrator option which combines a symplectic integrator
with a Bulirsch-Stoer scheme appropriate for when particles approach
any of the massive bodies in the simulation. We have 
used the hybrid option of the code to follow the movement of 
each particle for 30,000 years (using independent 
integrations for each test particle). Our choice of 30,000 years 
for the time that biological material can remain viable in space is 
adopted following Gladman (2005), who argue that unless the sensitivity 
of ancient microorganisms to radioactivity (the main killing factor in 
ejecta greater than a few meters) is 2-3 orders of magnitude greater 
than that of modern day bacteria, then survival times for viable biological 
material can range from 3,000 to 30,000 years.

The code stops the integration if the test particle collides with 
a planet or the Moon, if it reaches distances smaller than 1$R_{\odot}$ 
from the Sun, or if it is ejected from the simulation domain
(presently set at 40 AU). 
The base time step for the symplectic integrator is 24 hours, and 
it is decreased when the code switches to a Bulirsch-Stoer integrator
during close encounters between test particles and massive objects to 
achieve a given accuracy. In the present simulations the change from 
one integrator to another is set to occur when test particles are 
within 3 Hill radii of a massive body.

\subsection{Initial conditions}

We have analyzed several cases, each consisting of 10,242 particles,
set off with a given speed, $V_{\rm ej}$. Particles are distributed
uniformly over the surface of a sphere. To do so we use a Fuller
spherical distribution (Prenis 1988 and Saff and Kuijlaars 1997),
we begin with an icosahedron (by construction each of the 12 vertices of this
figure is on the circumscribed sphere). An icosahedron has 30 possible lines between each point
and its nearest neighbours. We take the middle point of each one of these lines and we project
them into the sphere. Therefore obtaining a new figure with 30+12=42 vertices uniformly
distributed on the sphere. The general formula to obtain the number of vertices $n_{p}$ is giving
by the following recursive formula:
\begin{equation}
n_{p}(j)=n_{lines}(j-1)+n_{p}(j-1)
\label{npart}
\end{equation}
Where $n_{p}(j)$ is the number of vertices in the step $j$, $n_{lines}(j-1)$ and $n_{p}(j-1)$ are
the number of lines and number of vertices in the previous step $(j-1)$, respectively. It is
important to notice that $n_{p}(1)=12$. It can be proven that $n_{lines}(j-1)=3 n_{p}(j-1) - 6$.
Therefore we can simplify Eq. \ref{npart} and obtain the following:
\begin{equation}
n_{p}(j)=[3 n_{p}(j-1) - 6] + n_{p}(j-1)=4  n_{p}(j-1) - 6
\label{npartsimpl}
\end{equation}
Therefore, using Eq. \ref{npartsimpl} we obtain 10,242 vertices after 5 iterations, we place our
particles on each one of these vertices. The points are
uniformly distributed on the sphere.


\begin{figure}[!t]
\includegraphics[width=\columnwidth]{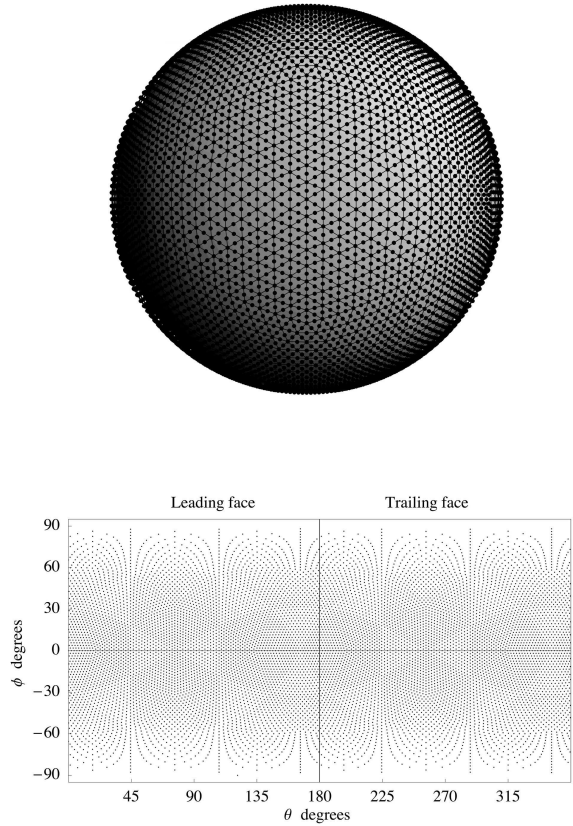}
\vspace{-1.3cm}
\caption {Initial spatial distribution of ejected particles at 
a height of 100 km over the surface of the planet. The bottom 
panel shows the initial location of particles as a function of 
latitude ($\phi$) and ``longitude'' ($\theta$ (measured from the 
midnight meridian).}
\label{fig1}
\end{figure}


The initial ephemeris of the planets-moon system is taken from 
the Horizons website ({\it http://ssd.jpl.nasa.gov/horizons}) 
and correspond to the configuration of the Solar System on the 
6th of July 1998 at 00:00:00.0 UT. This date is the default starting 
date defined in the Mercury 6.2 code examples, it does not represent any
special configuration of the planetary bodies and is adopted as 
an arbitrary initial condition. In section 4 we discuss the
effect of changing the initial ephemerides on some of our results. 
  
Test particles are set off from a height of 100 km over 
the surface of 
the Earth with an initial velocity that is purely radial. Of course,
this velocity distribution does not correspond to any particular 
impact, as these would most likely result in the preferential ejection 
from one side of the planet (over a single quadrant) and ejecta with 
a distribution of velocities. Rather, by choosing 
the ejection velocity in this manner, and analysing several cases 
with a single ejection speed separately, we intend to sample all likely 
ejection conditions and identify trends indicating the effect of 
the ejection parameters, speed, launch position and direction. 
In the discussion section, we analyse the 
dependence of the collision probability with different bodies on the
location from which a test particle is launched, although this can 
only be done with reasonable statistics for collisions with Earth 
and, to a lesser extent, with Venus and Jupiter.      

According to Gladman et al. (2005), in the impact spallation theory of 
Melosh (1985) the cumulative distribution of ejected mass 
from Earth, from the escape velocity up to a given ejection velocity,
is given by:

\begin{equation}
F(V_{\rm esc} < V < V_{\rm ej}) = 
 \frac{1-(\frac{V_{\rm ej}}{V_{\rm esc}})^{-5/3}}{1-(\frac{U}{2 V_{\rm esc}})^{-5/3}}
\end{equation}

\noindent where $F$ is the fraction of the total mass ejected that 
leaves Earth with velocity in the range $V_{\rm esc} < V < V_{\rm ej}$ and
$U$ is the impactor speed. Note that the distribution 
is unity when $V_{\rm ej} = U/2$, meaning that the maximum velocity 
with which material is ejected is one half of the 
speed with which the impactor collides with Earth. The distribution 
function for Earth-crossing asteroids or comets drops steeply as the
impactor velocity increases, so that ejecta with velocity much greater than
the escape velocity are even less likely to occur. The range of ejection 
velocities we consider correspond to impacts with speed less than 33 km/s, 
covering most of the asteroid and comet impacts with the Earth and 
Moon (Chyba et al. 1994). In the present paper we do not 
consider ejecta moving with even higher velocity
which may result from Earth impacts by objects moving on 
higher velocity, comet-like trajectories. 

\medskip

\begin{table*}[!t]\centering
  \caption{Description of the cases studied and summary of the 
	number of collisions with different bodies according to 
	our results. In parenthesis we give the percentage relative
	to the total number of bodies in the simulation, i.e. the 
	collision probability.} \label{tab:1}
 \begin{tabular}{lrrrrr}
    \hline
    Case    & A & B & C & D & E \\
    \hline
    $V_{\rm ej}$ (km/s) & 11.22 & 11.71 & 12.7 & 14.7 & 16.4 \\
    \hline
    Earth   & 496  & 106  & 48  & 22     & 10  \\
        & (4.84\%) & (1.03\%) & (0.47\%) & (0.21\%)    & (0.1\%) \\
    Moon    & 2   & 2 & 2  & 1 & 0 \\
         & (0.02\%)   & (0.02\%) & (0.02\%)  & (0.01\%) & (0\%) \\
    Venus   & 6   & 17    & 7  & 7 & 3 \\
        & (0.06\%)   & (0.17\%)    & (0.07\%)  & (0.07\%) & (0.03\%) \\
    Mars    & 0    & 1    & 1     & 0     & 0 \\
         & (0\%)      & (0.01\%)    & (0.01\%)     &  (0\%)    &  (0\%) \\
    Jupiter & 0       & 0     & 0      & 6     & 5  \\
      &  (0\%)      &  (0\%)    &  (0\%)     &  (0.06\%)    &  (0.05\%) \\
    Sun     & 0       & 0     & 0      & 0     & 19  \\
          &  (0\%)      &  (0\%)    &  (0\%)     &  (0\%)    &  (0.19\%) \\
    Ejected$^*$ & 0  & 0  & 0  & 254  & 691  \\
       &  (0\%) &  (0\%) &  (0\%) &  (2.48\%) &  (6.75\%) \\
    \hline
  \end{tabular}
\begin{flushleft}
 $^*$ \footnotesize{Particles reaching distances greater than 40 AU.}
\end{flushleft}
\end{table*}


\section{Results}

A series of numerical simulations with different ejection velocity were 
conducted, cases with low, intermediate and high ejection velocity 
are considered, corresponding to specific cases also reported in the 
study by Gladman et al. (2005). This allows a direct comparison to the 
results reported by these authors. We also report results for two 
additional cases taken at intermediate values of the ejection velocity, 
which do not exactly coincide with the cases studied by Gladman et 
al. (2005). Table 1 shows a summary of our results for each of these cases.


\begin{figure}[!t]
\includegraphics[width=\columnwidth,height=1.7\columnwidth]{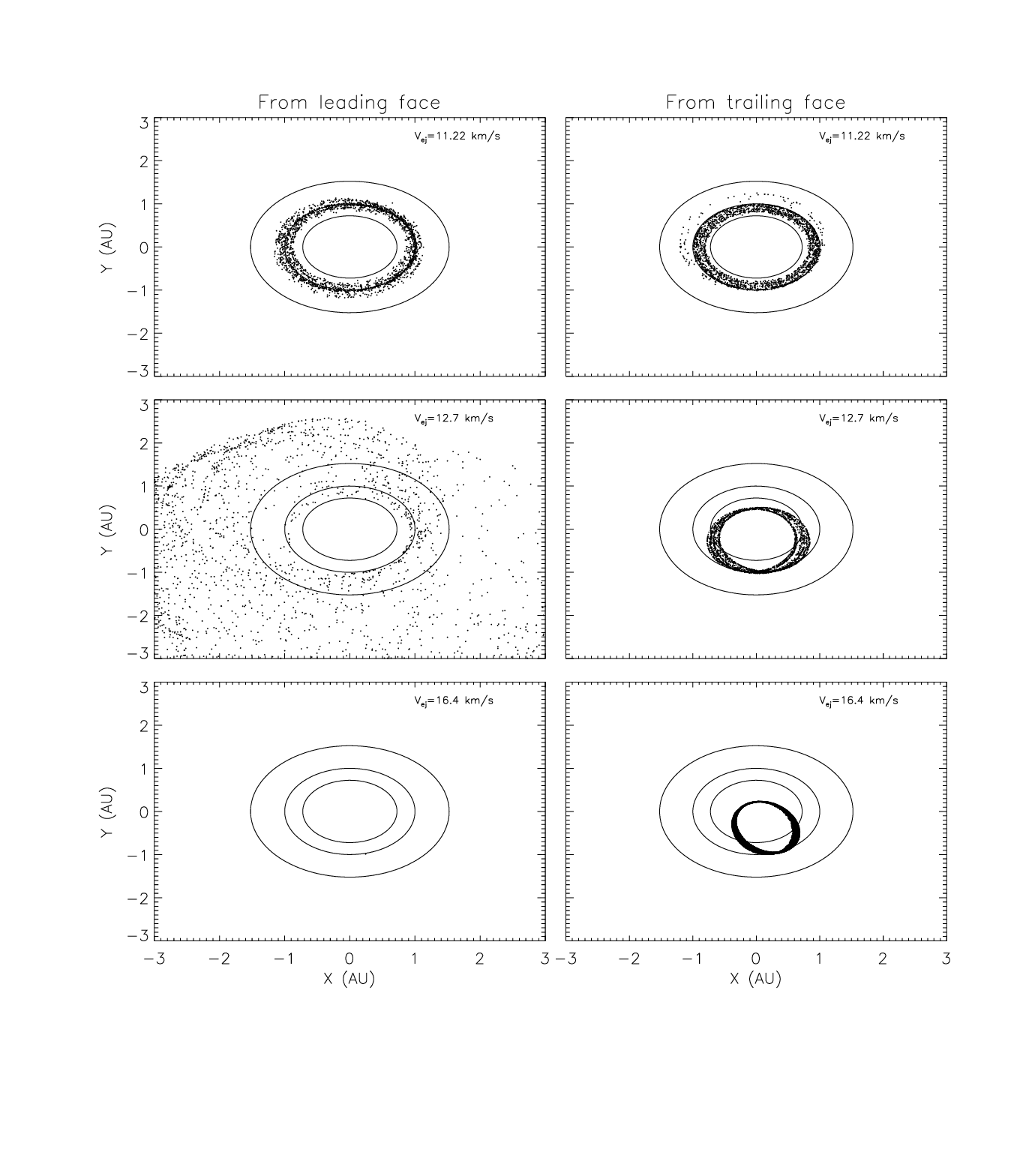}
\vspace{-2.5cm}
\caption{Projection in the $x-y$ plane of the trajectory of 
2 particles, one ejected from center of the leading face of the planet, 
along its direction of motion (left column of panels), and the other
ejected from the center of the trailing face. Dots in each panel 
denote the position of the particle at each of the output times of our 
simulations. Top, middle and bottom rows correspond to different 
ejection velocities, Cases A, C and E (Table 1), respectively.}
\label{fig2}
\end{figure}


Neglecting the effect of the other bodies in the simulation, the ejection 
velocity from Earth determines the maximum apoapsis and minimum 
periapsis in the orbit of the ejected particles around the Sun. 
The maximum apoapsis is reached by 
particles ejected from the leading face of the planet, where the net 
velocity with respect to the Sun is maximum. Oppositely, particles
moving in the direction opposing the motion of the planet at the moment 
of ejection, will have the smallest heliocentric velocity and fall to the
minimum periapsis.  Since the particles we are 
studying are launched very close to the Earth
we must take Earth's gravitational potential into account.
The velocity "at infinity" ($V_{\rm \infty}$) is defined as the speed 
that ejecta have after escaping  the planet's gravitational well 
and is related to the ejection speed ($V_{\rm ej}$) by the following:

\begin{equation}
V_{\rm \infty}^2=V_{\rm ej}^2-V_{\rm esc}^2 ,
\label{vinfty}
\end{equation}

\noindent where $V_{\rm esc}=\sqrt{2 G m_{E} / r_{E}}$,  is the escape 
velocity from the planet, $G$ is the gravitational constant, $m_{E}$ is 
the mass of the Earth and $r_{E}$ is the distance from the ejected particle 
initial position to the centre of the Earth. In the present study we 
adopt $r_E$ = 6471 km, so that $V_{\rm esc}$ = 11.098 km/s. 
In order to estimate the periapsis and apoapsis we 
assume that the particles have already escaped the gravitational potential 
of the Earth and have the velocity given by Eq. (\ref{vinfty}) but since 
they have been launched from Earth's surface we need to add the velocity 
of our planet, taken as the average along its orbit, $V_{E}=29.29$ km/s, 
so the particle velocity is given by:

\begin{equation}
V_{part}=V_{E}\pm V_{\rm \infty}
\label{vpart}
\end{equation}

In Eq. (\ref{vpart}) we use the positive sign if the particle is 
ejected in the leading face and negative otherwise.
Let us assume that the Earth is in a circular orbit, then the angular 
momentum of the particle is given by $h=r_{p} V_{part}$, where $r_p$ is
the position of the particle with respect to the Sun. At the 
time of ejection, $r_{p} \approx 1$ AU, but with a velocity that is too big 
for staying in a circular orbit (or too small for the negative sign
in Eq. (\ref{vpart})), the new semi-major axis and 
eccentricity of the ejected particle can be calculated using Eqs. (2.134) 
and (2.135) from Murray and Dermott (1999):

\begin{equation}
a=\Bigg( {2\over r_{p}} - \frac{V_{part}^2}{G M_{\odot}} \Bigg)
\label{apart}
\end{equation}

\begin{equation}
e=\sqrt{ 1 - \frac{h^2}{G M_{\odot} a} }
\label{epart}
\end{equation}

\noindent where $a$, and $e$ are the semi-major axis and eccentricity of 
the new orbit of the particle, and $M_{\odot}$ is the mass of the Sun. 
Then for a given velocity of the particle $V_{part}$ we can have the 
following maximum apoapsis (taking positive sign in Eq. (\ref{vpart})) 
and minimum periapsis (taking negative sign in Eq. (\ref{vpart})):

\begin{equation}
Q_{max}=a_{max} (1 + e_{max}) 
\label{apomax}
\end{equation}

\begin{equation}
q_{min}=a_{min} (1 - e_{min}) 
\label{perimin}
\end{equation}

Where $a_{max}$ and $e_{max}$ are calculated from Eqs. (\ref{apart}) 
and (\ref{epart}) using the positive sign for $V_{part}$ and similarly 
$a_{min}$ and $e_{min}$ are calculated using the negative sign. According 
to these formulas, An ejection velocity of 11.62 km/s 
is needed to reach Mars and 14.28 km/s are required to reach the orbit of Jupiter. 

Figure 2 illustrates typical trajectories
followed by particles in the low, intermediate and high ejection 
velocity cases (A, C and E in Table 1). Shown are
the orbits of 2 particles during the 30,000 years integration for each case, 
one ejected from the central regions of the leading face of 
the planet (along its direction of motion) and the other 
from the center of the trailing face. These cases represent extrema 
in the orbital energy of ejected particles and are illustrate the
range of possible trajectories. Dots in each panel of Figure 2
denote the position of the particle in the $x-y$ plane at each of the 
output times of our simulations, separated by $\Delta t = 5000$ days. 

Particles ejected with the lowest velocity considered, $V_{\rm ej} = 11.22$ km/s,
generally remain in orbits close to that of the Earth, as shown
in the top panels of Figure 2. Those ejected with the highest velocity, 
$V_{\rm ej} = 16.4$ km/s (shown in the bottom panels of Figure 2), have 
access to a wide range of orbits. In this case, many particles, 
as the one depicted in the bottom left panel of Figure 2, 
are launched to the periphery of the Solar System and spend a very short 
time in the inner Solar System, hence the absence of dots in the figure.     
 
\subsection{Collision probability}

As indicated in Table 1, the probability that particles collide with 
Earth or any other body of the Solar System, depends strongly on the
velocity with which it is ejected from Earth. Particles ejected with
a low velocity never cross the orbit of Mars or Jupiter, and they 
can only collide with Venus, the Moon or fall back to Earth. 

In the context of our calculations, particles ejected with a velocity  
just 1\% greater than the escape velocity, $V_{\rm ej} = 11.22$ km/s 
(Case A in Table 1) have a maximum probability of falling back to 
Earth. Within 30,000 years, almost 5\% of all particles fall back to 
Earth. Particles do not have enough energy to reach the orbits of Mars, 
as exemplified in the top panel of Figure 2 and hence, there 
are no collisions with Mars and Jupiter. There are two 
particles that impact the Moon, and six that impact Venus, representing
0.02 and 0.06\% of the ejected population of test particles, respectively. 

In Case B of Table 1, characterized by ejection velocity  
$V_{\rm ej} = 11.71$ km/s, there are 106 particles that have fallen back
to Earth (1.03\% of the 10242 ejected particles) by the end of the 30,000 yr
simulations, 2 that hit the 
Moon (0.02\%) and 17 collide with Venus (0.17\%). 
In contrast to case A, in case B ejected particles now have 
enough energy to reach Mars, and one particle collides with the planet, 
representing 0.01\% of the ejecta. Ejection energy is still not enough 
to reach any of the outer planets.     

As the ejection velocity increases to $V_{\rm ej} = 12.7$ km/s, Case C,
the number of particles falling back to Earth continues decreasing, 
in comparison to cases A and B, with only 0.47\% of the 10242 particles
returning to Earth. As in Case B, 2 particles impact the Moon,
7 particles collide with the planet Venus and 1 particle reaches Mars. 

Of the particles ejected with conditions of case D, $V_{\rm ej} = 14.7$ km/s, 
only 0.21 \% return to Earth, continuing the trend observed in the previous 
cases. One particle impacts the Moon and 7 collide with Venus. Ejecta
now can have enough energy to reach the orbit of Jupiter and many particles 
do so. As a result, 6 particles collide with this planet representing 0.06 \% 
of the total number of ejected particles. The high velocity and low gravitational 
pull of Mars result in the absence of collisions with Mars for this case.
Also, a significant amount of 
particles are launched into orbits reaching the periphery of the Solar System,
254 particles travel beyond 40 AU. Purely for numerical efficiency reasons,
we consider these particles as ejected from the Solar System. 

Finally, only 0.1 \% of the particles ejected with the maximum velocity we 
considered, $V_{\rm ej} = 16.4$ km/s, ever return to Earth, no particle 
impacts the Moon or Mars, and only 3 particles impact Venus. A great number of 
particles reach the outer planets region, 5 of them colliding with Jupiter and 
691 ``escaping'' from the Solar System (as defined above). The low orbital 
velocity of particles ejected from the trailing face of the planet results 
in about 0.2\% of particles colliding with the Sun. Similarly to the 
previous case, no particles colliding with the planet Mars are found.      


\begin{figure}[!t]
\includegraphics[width=\columnwidth]{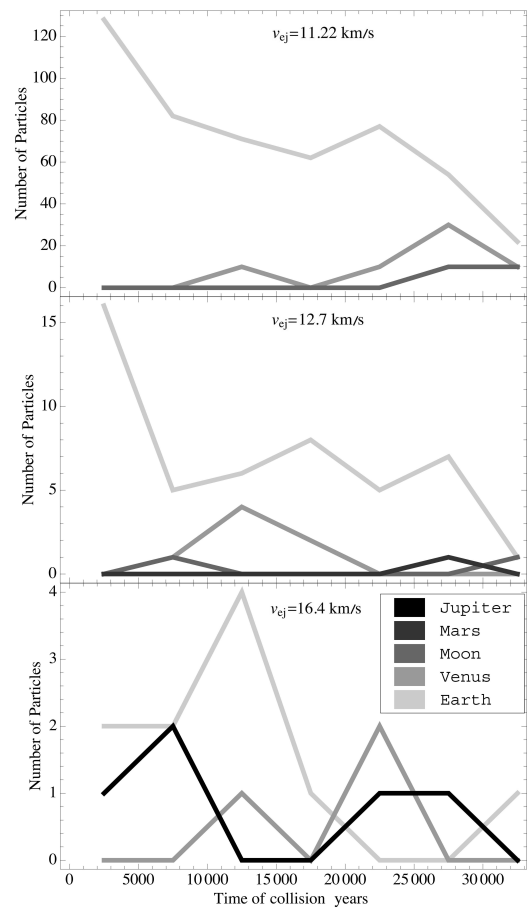}
\caption {Collision time for particles ejected with $V_{\rm ej} = 11.22$ km/s,
binned into 5000 yr intervals. Shades of gray correspond to particles 
falling back to Earth, colliding with the Moon, Venus Mars of Jupiter, 
as indicated in the Figure legend. Note that to make the plot clear,
the values corresponding to collisions with Venus and the Moon in 
the top panel are 10 times their real value.}
\label{fig3}
\end{figure}


\subsection{Collision time}

By collision time we mean the period a particle spends in space until
it collides with the Earth or any other Solar System body. In Figure 3,
we illustrate the collision time for particles falling back to Earth,
colliding with the Moon, with Venus, with Mars or with Jupiter, for
different values of the ejection velocity (Cases A, C and E).
Results are binned into 5000 year intervals with each line indicating
collisions with a different body as indicated in the figure legend.
It must be pointed out that in general, the number of collisions with
bodies other than Earth is very small, and the trends in the time evolution
of collision rates with such bodies, are of questionable
statistical significance. Further studies with a greater number of ejected
particles, necessary to address this issue in greater depth, are beyond
the scope of the present paper.

The top panel of Figure 3 corresponds to Case A, with
$V_{\rm ej} = 11.22$ km/s. In this case, more than 2/3 of the particles
returning to Earth do so in less than 20,000 years. The number of
fallback particles does not decrease strictly monotonically, but
less than 10\% of particles return in the last time bin, between 25 and
30 kyr after ejection. Also shown in the top panel, are the collision times
for particles colliding with the Moon and Venus. The numbers shown in the
figure are 10 times the actual values, which are too small to be plotted.
A few of the particles colliding with Venus do so after less than 15 kyr
but the majority take more than 20 kyr to reach the planet.

The middle panel of Figure 3 shows the collision times for particles ejected
with $V_{\rm ej} = 12.7$ km/s, Case B discussed above. The number of particles
falling back to Earth peaks at early times, approximately 15\% of particles
return in the first 5000 years. The fall-back rate more or less remains constant
after this and for the next 25 kyr. Only two particles collide with the Moon
and they do so at widely different times, the first one after less than 10 kyr
and the second one at the very last time bin, after 25 kyr. The single particle
that hits Mars does so near the end of our simulation, after 20 kyr. All
particles travelling to Venus do so in less than 20 kyr, with the peak in the
collision time to the planet between 10 and 15 kyr.

Finally, collision times for particles ejected from Earth with $V_{\rm ej} = 16.4$ km/s
are shown in the bottom panel of Figure 3. Only a few particles fall-back
to Earth, the peak in the number of these is between 10 and 15 kyr and by 20 kyr,
90\% of the particles that do so, have returned to Earth. Only 3 particles
collide with Venus, 1 of these impacts after less than 15 kyr and the rest
do so before 25 kyr. Of the 5 particles reaching Jupiter, a group representing
40\% of the colliders do so within just 10 kyr after being ejected, the remaining
60\% collide with Jupiter towards the end of the simulation. 

\section{Discussion}

Gladman et al. (2005) has performed a similar calculation to the one 
conducted in this paper using a different numerical code, initial conditions 
and a smaller number of test particles in each simulation. In general, 
our results agree well with those they report. Two notable exceptions,
most likely attributable to the greater number of test particles we 
follow in our simulations, are that we find collisions with Mars, one 
particle in Cases B and C, and also, we find collisions with Jupiter, 
0.06\% of all ejecta in Case D and 0.05\% in Case E. Using an 
\"Opik collision probability calculation, Gladman et al. (2005) 
estimated the collision rate with Mars to be about 2 orders of magnitude 
lower that found on the basis of our simulations. However, as also noted 
in their paper, our results  for Mars are within the known typical errors 
of such probability estimations. No collisions with Jupiter are reported 
in Gladman et al. (2005).

\begin{figure}[!t]
\includegraphics[width=\columnwidth,height=1.33\columnwidth]{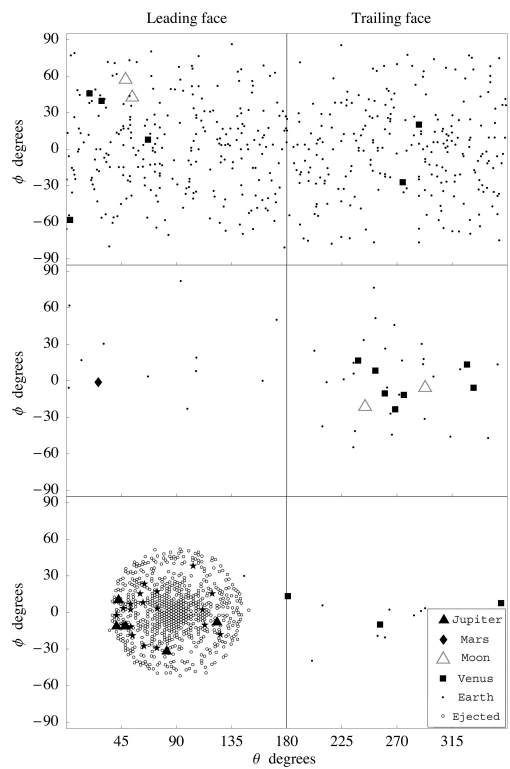}
\caption {Initial location of particles that collide with different
bodies in our simulation. Angle $\phi$ is the latitude of the launch
position for each particle, $\theta$ is a longitude like angle but
measured from the midnight meridian. The top panel corresponds to 
colliding particles ejected at low velocity (Case A), the middle panel 
corresponds to intermediate ejection velocity (Case C) and the bottom 
panel shows the initial location of colliding particles in the highest ejection
velocity considered (Case E). The figure legend at the bottom right corner
indicates the body with which a particle collides.}
\label{fig4}
\end{figure}

Both results, definite collisions with Mars and Jupiter,  
are of astrobiological significance, owing to the possible 
presence of life sustaining environments in early Mars and in Jupiter's moons 
Europa and Ganymedes. Also worth noting is the fact that the 
single particle colliding with Mars, 
does so towards the end of our simulation, between 25 and 30 thousand years 
after being ejected from Earth. Collisions with Jupiter are characterized by 
a wider range of collision times, one half reaching the giant planet in less 
than 10,000 years. In future studies we will extend our analysis of both cases 
to determine the statistical significance of these results. 

\subsection{Effect of the Moon}

In order to estimate the effect of the presence of the Moon on the 
dynamics of ejecta, we have performed a simulation with exactly the 
same initial conditions and run parameters as Case D (Table 1), but,
as is done in Gladman et al. (2005), assuming that the Earth and the Moon 
are integrated into a single body with the combined mass and located 
at the center of mass of the system. Only minor differences in 
the results are found, in comparison to Case D with the Earth and the Moon 
as separate bodies. In the single body Earth-Moon simulation, 23 particles 
fallback to the Earth-Moon (22 in Case D), 8 particles impact Venus (7 in Case D), 
4 collide with Jupiter (6 in Case D) and 2 reach the Sun (0 in Case D). 
A small difference is also found in the number of particles ``ejected'' 
from the system, 220 in this case (254 in Case D).

\subsection{Effect of ejection location}

In Figure 4, we plot the distribution of launch positions of fall-back and
colliding particles. Each panel depicts the initial latitude (angle $\phi$)
and a longitude-like angle ($\theta$ is measured from west to east but starting
from the midnight meridian at the time of ejection). For example, the point
$\phi = 0$ and $\theta = 90^{\rm o}$ corresponds approximately to the center
of the leading face along the direction of motion of the planet.

The top, middle and bottom rows of Figure 4, show the initial location 
of particles ejected with $V_{\rm ej} = 11.22$, $V_{\rm ej} = 12.7$ and
$V_{\rm ej} = 16.4$ km/s, Cases A, C and E, respectively. No clear asymmetry 
is found in the distribution of ejection locations for particles falling back 
to Earth in the case with $V_{\rm ej} = 11.22$ km/s. This can be understood from 
the fact that ejecta with such a low velocity remain in orbits which are 
very close to Earth, as illustrated in Figure 2, thus increasing the chance of 
collision. As expected, there is also no asymmetry with respect to the equator. 
The slight asymmetries in the ejection location for particles eventually 
colliding with Venus and the Moon, are probably not statistically significant, and
must be tested in future simulations with an even greater number of particles.

In the case with intermediate ejection velocity, $V_{\rm ej} = 12.7$ km/s, 
shown in the middle panel of Figure 4,               
the number of particles falling back to Earth is slightly higher for those 
ejected from the trailing face, since orbits of these particles are less dispersed,
i.e. more concentrated in the vicinity of the Earth, than for particles 
ejected from the leading face (see Figure 2). For the same reason, we also find that particles 
ejected from the trailing face are more likely to impact Venus. The opposite is true
for particles traveling to Mars, as ejecta from the 
trailing face in this velocity range do not have enough energy to reach the planet.

The highest ejection velocity we consider, $V_{\rm ej} = 16.4$ km/s, leads to 
ejecta capable of traveling outside the planetary region of the Solar System.
We label these particles as ejected since they spend a very short amount of time 
in the inner Solar System, so that their collision probability with other planets is
negligible. These are shown in the bottom panel of Figure 4 and they arise 
exclusively from the leading face of the planet. A similar asymmetry in 
the ejection location is found for particles colliding with Jupiter, since 
only particles ejected with a high total velocity are capable of reaching 
the planet. A few particles are found to fall-back to Earth and to collide with 
Venus, mostly ejected from the trailing face.  
    
These results suggest that the probability of collision  with different
Solar System bodies of Earth ejecta resulting from a giant impact, is clearly 
dependent on the particular place on Earth where the collision occurred. 
Impacts on the leading face of the planet along its direction of motion,
which are statistically more likely, lead to ejecta that have a higher probability
of colliding with Mars and Jupiter. 

\subsection{Effect of initial ephemeris}

We have also performed a simulation with the same 
ejection velocity as Case A, but using a different initial ephemerides 
for the planets, corresponding to an initial time 6 months
after the start of the rest of the simulations reported in this
paper. This allows us to determine whether the
initial planet configuration has a significant effect on the 
dynamics of ejecta and the resulting collision probabilities. Only 
small differences in our results are found in comparison to Case A.
In the case of an initial ephemerides taken 6 months later than that
considered in Case A, 531 particles fallback to Earth (496 in Case A), 
8 particles impact Venus (6 in Case A) and 1 reaches the 
Moon (2 in Case A). A difference of 7\% in the number of particles 
returning to Earth is found as a result of the different initial 
planetary configuration. The statistical significance of the differences in the
number of collisions with the Moon and Venus, must be verified in 
future simulations with a greater number of ejected particles. 

\section{Conclusions}

We have computed the trajectory of an ensemble of particles 
representing ejecta from Earth, resulting from the giant impact 
of a comet or asteroid, in order to determine the collision probability 
with different Solar System bodies. Several ejection velocities, 
representing the different ejecta components in a given impact, as 
well as different initial planetary configurations, have been explored
with simulations over a period of 30,000 years.
In agreement with previous work (Gladman et al. 2005) we find that 
ejecta can collide with the Moon, Venus or fallback to Earth after
a period of several thousand years in space. A novel result in our 
simulations is finding particles that collide with the planet Mars,
for intermediate ejection velocities, and also with Jupiter, for 
high ejection velocity.  

Of course, a given impact will give rise to ejecta with a wide
spectrum of velocities, the maximum determined by
the speed of the impactor as it hits the Earth. This, together
with other characteristics of the impact, also defines
the amount of material ejected to space. In general, 
most of the escaping material does so with a velocity slightly 
greater than the Earth's escape velocity, the number of 
particles ejected drops rapidly as ejection velocity increases
(Eq. (1)). Hence, the calculation of the net collision probability with a 
given Solar System body, must take into account the sharply 
decreasing velocity distribution of ejecta.
 
On the basis of our results, and considering the rule of thumb that the 
maximum ejection velocity is one half of the impactor speed, we must
conclude that: 1) In collisions with impactor speed greater than 
2 $V_{\rm esc}$, a significant amount of material, of the order of a few percent,
will fall back to Earth after remaining in interplanetary space for
less than 30 kyr, 2) Ejecta transfer to Venus and the Moon can occur 
as long as $U \gtrsim 2 V_{\rm esc}$, 3) Ejecta transfer 
to Mars requires an ejection speed only slightly greater, less than 5\% more,
than the escape velocity and 4) The transport of terrestrial
crustal material to the vicinity of Jupiter requires 
an impactor speed of almost 3 times the escape speed, and can 
occur only if the impact is on the leading face of the planet
as it orbits around the Sun.

A more detailed calculation of the collision probability, taking 
into account the velocity distribution of ejecta, as well
as computations with a greater number of particles to estimate 
statistically significant collision rates with the Moon, Mars and Jupiter, 
will be the subject of future contributions.   

\bigskip

\noindent {\bf{Acknowledgements}}

\smallskip

The authors acknowledge support from research grants IN109409 
of DGAPA-UNAM and CONACYT-M\'exico grant No.128563.

\end{document}